\documentstyle[epsfig,psfig]{article}

\title{MAXIPOL: A Balloon-borne Experiment for Measuring the
Polarization Anisotropy of the Cosmic Microwave Background Radiation}

\author{B. R. Johnson$^{1}$, 
M. E. Abroe$^{1}$, P. Ade$^{3}$, J. Bock$^{4}$, J. Borrill$^{7,9}$,\\
J. S. Collins$^{2}$, P. Ferreira$^{5}$, S. Hanany$^{1}$, A. H. Jaffe$^{8}$,\\
T. Jones$^{1}$, A. T. Lee$^{2,6}$, L. Levinson$^{10}$, T. Matsumura$^{1}$, B. Rabii$^{2}$, T. Renbarger$^{1}$,\\
P. L. Richards$^{2}$, G. F. Smoot$^{2,6}$, R. Stompor$^{7}$, H. T. Tran$^{2}$, C. D. Winant$^{2}$\\
\\
{\small $^{1}${\it School of Physics and Astronomy, University of Minnesota, Minneapolis, MN, USA}}\\
{\small $^{2}${\it Department of Physics, University of California, Berkeley, CA, USA}}\\
{\small $^{3}${\it Department of Physics and Astronomy, University of Wales, Cardiff, UK}}\\
{\small $^{4}${\it Jet Propulsion Laboratory, Pasadena, CA, USA}}\\
{\small $^{5}${\it Astrophysics \& Theoretical Physics, University of Oxford, Oxford, UK}}\\
{\small $^{6}${\it Physics Division, Lawrence Berkeley National Lab, Berkeley, CA, USA}}\\
{\small $^{7}${\it Computational Research Division, Lawrence Berkeley National Lab, Berkeley, CA, USA}}\\
{\small $^{8}${\it Astrophysics Group, Blackett Lab, Imperial College, London, UK}}\\
{\small $^{9}${\it Space Sciences Laboratory, University of California, Berkeley, CA, USA}}\\
{\small $^{10}${\it Department of Particle Physics, Weizmann Institute of Science, Rehovot, Israel}}
}

\begin{document}

\maketitle
\bibliographystyle{plain}

\abstract{We discuss MAXIPOL, a bolometric, balloon-borne experiment
designed to measure the E-mode polarization anisotropy of the cosmic
microwave background radiation (CMB) on angular scales of 10$^\prime$
to 2$^{\circ}$. MAXIPOL is the first CMB experiment to collect data
with a polarimeter that utilizes a rotating half-wave plate and fixed
wire-grid polarizer.  We present the instrument design, elaborate on
the polarimeter strategy and show the instrument performance during
flight with some time domain data.  Our primary data set was collected
during a 26 hour turnaround flight that was launched from the National
Scientific Ballooning Facility in Ft. Sumner, New Mexico in May 2003.
During this flight five regions of the sky were mapped.  Data analysis
is in progress.}


\section{Introduction}
\label{sec:introduction}

MAXIPOL is a bolometric, balloon-borne experiment designed to measure
the E-mode polarization anisotropy in the cosmic microwave background
radiation (CMB). The MAXIPOL instrument is a reimplementation of the
hardware from the successful CMB temperature anisotropy experiment
MAXIMA \cite{hanany,lee,stompor}. While the MAXIMA telescope and data
electronics remained largely unchanged, the receiver was converted
into a polarimeter by retrofitting it with a rotating half-wave plate
(HWP) and a fixed wire-grid polarizer.

MAXIPOL has flown twice from NASA's National Scientific Ballooning
Facility in Ft. Sumner, New Mexico. The first flight, MAXIPOL-0,
launched in September 2002 and the second, MAXIPOL-1, in May 2003. In
this paper we discuss the science goals of the experiment, the
hardware implementation (Section~\ref{sec:instrument}), the flights
(Section~\ref{sec:flights}), the noise and potential systematic errors
(Section~\ref{sec:systematics}).

The goal of MAXIPOL is to measure the peaks in the E-mode (EE) and
temperature-E-mode (TE) cross correlation power spectra between
$\ell$=300 and $\ell$=1000. To accomplish this goal, MAXIPOL mapped
the $I$, $Q$ and $U$ Stokes parameters of 2$^{\circ}$ wide ``bow tie''
shaped regions of the sky with 10$^\prime$ resolution. Detection of
the polarization anisotropy of the CMB was recently reported by DASI
\cite{dasi} and WMAP \cite{wmap}.


\section{Instrument Description}
\label{sec:instrument}

Many subsystems in the MAXIPOL instrument have already been thoroughly
detailed in previous MAXIMA publications
\cite{hanany,lee_hardware,winant,rabii}.  This discussion will focus
primarily on the new MAXIPOL-specific hardware elements that were
retrofitted into the MAXIMA instrument.

\subsection{Overview}
\label{sec:overview}

\begin{figure}[t]
\vspace{-0.75in}
\centerline{\psfig{file=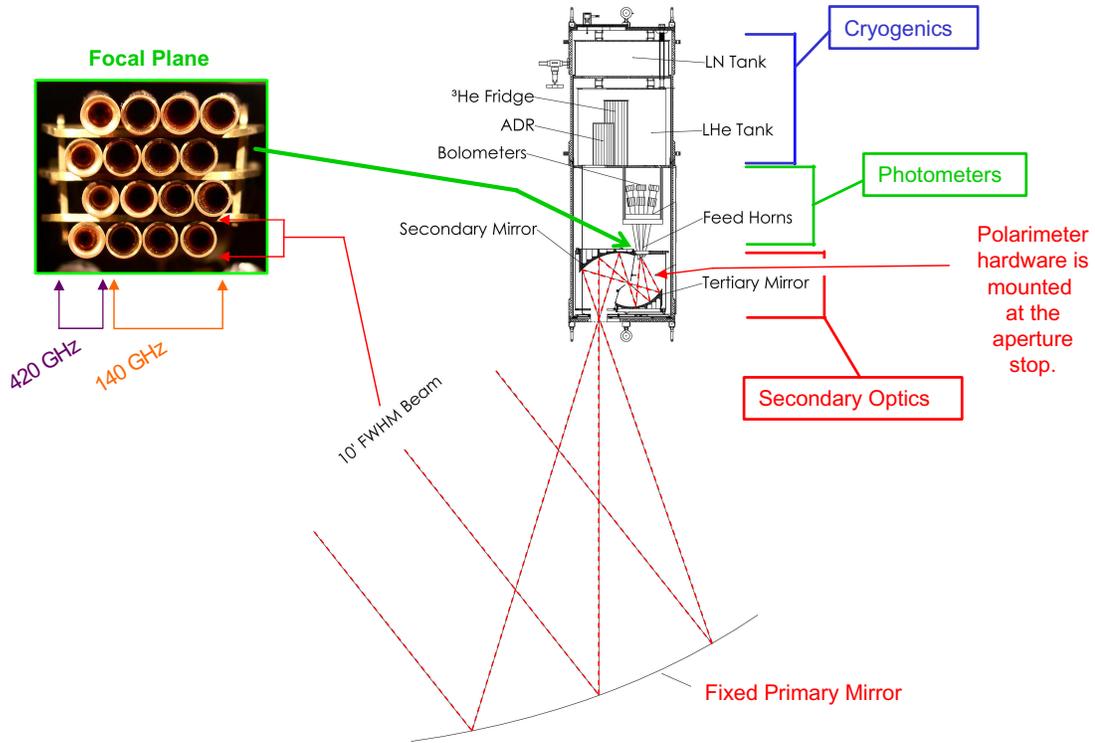,height=8.5in,angle=-90,clip=}}
\vspace{-0.5in}
\begin{center}
\parbox{6.5in}{
\caption{ \footnotesize A cross-sectional view of the MAXIPOL receiver
(Section~\ref{sec:instrument}).}
\label{fig:cryostat}
}
\end{center}
\end{figure}

MAXIPOL employs a three mirror, f/1 Gregorian telescope with a 1.3 m
off-axis parabolic primary mirror. The elliptical secondary and
tertiary reimaging mirrors (21 and 18 cm in diameter, respectively)
are held at liquid helium temperatures inside the receiver to reduce
radiative loading on the bolometers. To keep the instrumental
polarization properties of the telescope constant, all telescope
mirrors are fixed with respect to each other for all
observations.\footnote{The primary mirror was chopped in azimuth
during MAXIMA observations.}

Light from the sky is reimaged to a 4x4 array of horns at the focal
plane.  Observations are made in bands centered on 140 GHz and 420
GHz ($\Delta\nu \simeq$ 30 GHz).  The twelve 140 GHz photometers are
optimized to measure the CMB and the four 420 GHz photometers are
used to monitor foreground dust contamination. The 10$^\prime$ FWHM
Gaussian beam shape for the 140 GHz photometers is defined by a
smooth walled, single-mode conical horn and a cold Lyot stop; the 420
GHz photometers employ multi-mode Winston horns.  The bolometers
are maintained at 100 mK by an adiabatic demagnetization refrigerator
\cite{hagman} and a 300 mK $^{3}$He refrigerator.  A photograph of the
focal plane and a cross-sectional view of the inside of the cryostat
is shown in Figure~\ref{fig:cryostat}. For reference, the HWP and
wire-grid polarizer that will be discussed in
Section~\ref{sec:polarimeter} are mounted near the Lyot stop and at
the focal plane, respectively.

The telescope was focused before flight by mapping the detector beams
with a chopped, 100 Watt halogen filament imaged at infinity by a 38
inch on-axis parabolic mirror. An absorptive, 0.75 inch thick plug of
Eccosorb MF110 was inserted in the optical path at the intermediate
focus of the telescope to attenuate the intensity of warm loads during
lab measurements. This attenuator was anti-reflection (AR) coated with
a 0.015 inch thick sheet of etched Teflon. The calculated transmission
was $\sim$1\% at 140 GHz \cite{eccosorb}.

\begin{figure*}[t]
\vspace{-0.6in}
\centerline{\psfig{file=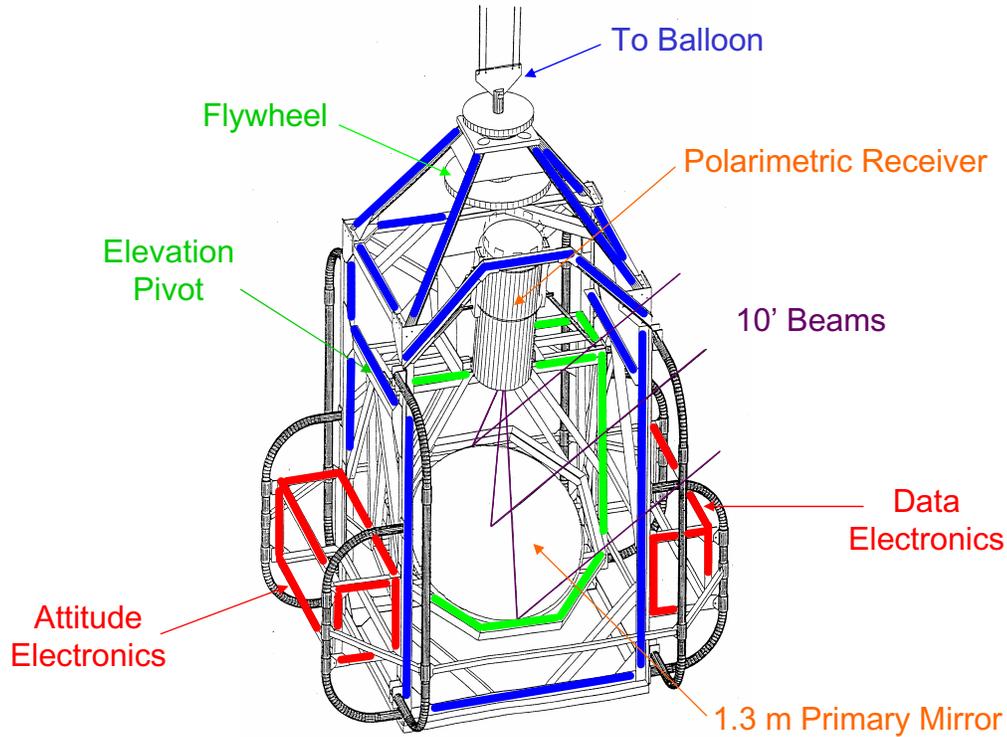,height=7.in,angle=-90,clip=}}
\vspace{-0.5in}
\begin{center}
\parbox{6.5in}{
\caption{ \footnotesize The MAXIPOL instrument without baffling
(Section~\ref{sec:instrument}).}
\label{fig:gondola}
}
\end{center}
\end{figure*}

A schematic of the MAXIPOL instrument can be seen in
Figure~\ref{fig:gondola}.  This illustration shows the payload without
sun shielding so the telescope, receiver and attitude control
subsystems are visible.  Before flight, sun shielding was installed to
protect all subsystems from solar radiation during daytime
observations and to shield the telescope and receiver from spurious
signals caused by sunlight and RF transmitters. The baffling was made
of Celotex aluminized foam sheeting and was painted white on all sun
and earth-facing surfaces.  We selected a white paint pigmented with
TiO$_{2}$ because this material has low solar absorptivity
($\sim$10\%) and high infrared emissivity ($\sim$90\%) -- a
combination that compensated for the loss of convective cooling in the
low-pressure balloon environment by providing adequate radiative
cooling.  This sun shield design successfully maintained all
instrument subsystems within nominal temperature specifications during
the daytime portion of the flight. A large aluminum ground shield
(also not illustrated) was mounted to the inner frame to shield the
main beam of the telescope from terrestrial emission.

Telescope attitude was feedback controlled. The flywheel mounted at
the top of the gondola moved the telescope in azimuth while the
elevation angle was adjusted with a linear actuator arm that nodded
the entire inner frame. A second motor mounted at the very top of the
gondola further assisted in moving the telescope in azimuth by
torquing the payload against the balloon cabling.  The azimuth
feedback-loop relied on gyros and a magnetometer and the elevation
feedback relied on a 16-bit optical encoder. The magnetometer was
calibrated before flight; the offset was measured to within a degree
and the non-linearity was mapped and stored in a lookup table that was
used by the on-board pointing computer during flight for making fine
corrections. A second lookup table was also implemented to account for
variations in the magnetic field of the Earth as a function of
longitude and latitude.

Pointing reconstruction for data analysis relies on the position of a
reference star in one of two boresight Cohu 4910 CCD cameras.  The
camera used during daytime observations was filtered with a 695
nm Schott glass filter and fitted with a 500 mm Promaster Spectrum 7
reflective lens that provided a 0.72$^{\circ}$ by 0.55$^{\circ}$
field-of-view; the unfiltered nighttime camera used a 50 mm Fujinon
lens that provided a 7.17$^{\circ}$ by 5.50$^{\circ}$ field-of-view.
Pixel size for the daytime and nighttime cameras was 0.084$^\prime$ by
0.069$^\prime$ and 0.84$^\prime$ by 0.69$^\prime$, respectively. The
small field-of-view and the filter on the daytime camera were
necessary to improve the ratio of star to sky brightness.  With the
combination, we detected stars of apparent visual magnitude 2 at
balloon altitude.  The two cameras and the telescope were aligned
before flight to within a quarter of a degree.

Bolometer data and housekeeping signals were multiplexed into a single
data stream that was telemetered to fixed ground stations during
flight. These signals were monitored in real time to ensure nominal
operation of the instrument and because the cryogenic system needed to
be manually cycled.

A new on-board data recorder was added to the experiment for the
MAXIPOL-1 flight after a NASA data transmitter failed during
MAXIPOL-0.  The data recorder, which was custom designed and built by
the Weizmann Institute of Science in Israel, consists of NIM modules
each containing an Altera FPGA chip and up to 128 Intel 16 MB flash
memory chips. With an uncompressed serial data rate of 160 kbps each
module is capable of storing 28.4 hours of data.  Individual modules
can be daisy-chained to each other to increase the total recording
capacity.  The FPGA chip reads the incoming data stream and when it
detects a pre-programmed frame structure it controls the storage of
the data on the memory chips. It also controls the post-flight export
of the data from the memory chips into a computer through a standard
parallel port.  Power consumption during steady-state data recording
is less than 0.25 Watt (at 5 V), and modules that are idle require
only about 0.05 Watt.  For MAXIPOL-1 two modules containing 96 memory
chips provided a total recording capacity of 42.6 hours. Approximately
28 hours of pre-flight, ascent and at-float data were recorded.

\subsection{Polarimeter Strategy}
\label{sec:polarimeter}

\begin{figure*}[t]
\centerline{\psfig{file=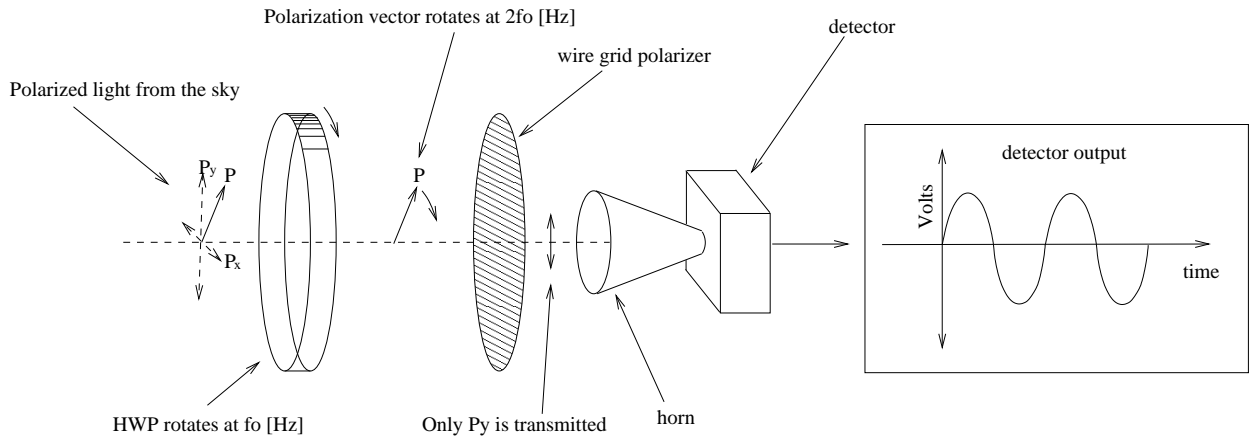,height=6.5in,angle=-90}}
\vspace{0.25in}
\begin{center}
\parbox{6.5in}{
\caption{ \footnotesize The polarimeter strategy employed by MAXIPOL
(Section~\ref{sec:polarimeter}).}
\label{fig:hwp}
}
\end{center}
\end{figure*}

MAXIPOL analyzed the polarization of the millimeter-wave sky with a
rotating HWP and fixed wire-grid polarizer. While this technique is a
well-known standard in astronomy, it is the first implementation in a
CMB experiment. The strategy is illustrated in Figure~\ref{fig:hwp}.

Monochromatic linearly polarized light that passes through a HWP
rotating at a frequency $f_{0}$ emerges linearly polarized with its
orientation rotating at $2f_{0}$. If this light then propagates
through a fixed polarizer and its intensity is subsequently measured,
the resulting data stream will exhibit sinusoidal modulation at
$4f_{0}$. The amplitude of this modulation depends on the level of
polarization of the incident radiation.  Perfectly polarized light
will maximize the amplitude and perfectly unpolarized light will yield
no modulation.

The advantage of HWP polarimetry is that each detector in the array
makes an independent measurement of the Stokes parameters of the
incoming radiation. In addition, this technique rejects systematic
errors. Spatial polarization variations on the sky translate to
temporal amplitude variations in the $4f_{0}$ signal because the
telescope is scanning.  Therefore, the polarization anisotropy data
will reside in the sidebands of the $4f_{0}$ signal in Fourier
space. Any spurious or systematic signals appearing in the data stream
outside of this $4f_{0}$ frequency band can be filtered away with
software during data analysis (see Section~\ref{sec:systematics}).

The HWP is inherently a monochromatic device so the behavior described
above applies only to the frequencies $\nu=mc/2t\Delta n$ where
$\Delta n$ is the difference between the ordinary and extraordinary
index of refraction in the birefringent crystal, $t$ is the
propagation length through the crystal, $m$ is an odd integer and $c$
is the speed of light.  Linearly polarized light at other frequencies
emerges from the crystal elliptically polarized.  We calculated the
HWP thickness that would minimize the fraction of elliptically
polarized intensity and thereby optimize the overall polarimeter
efficiency. To do this, we found the maximum of the product of the
expected efficiencies for the 140 and 420 GHz photometers as a
function of crystal thickness.  These expected efficiency curves
incorporated the spectral breadth of the photometers and the
convergence of rays as they propagate through the HWP. The HWP design
that resulted from this calculation is discussed in
Section~\ref{sec:hwp}.

\begin{figure*}[t]
\centerline{\psfig{file=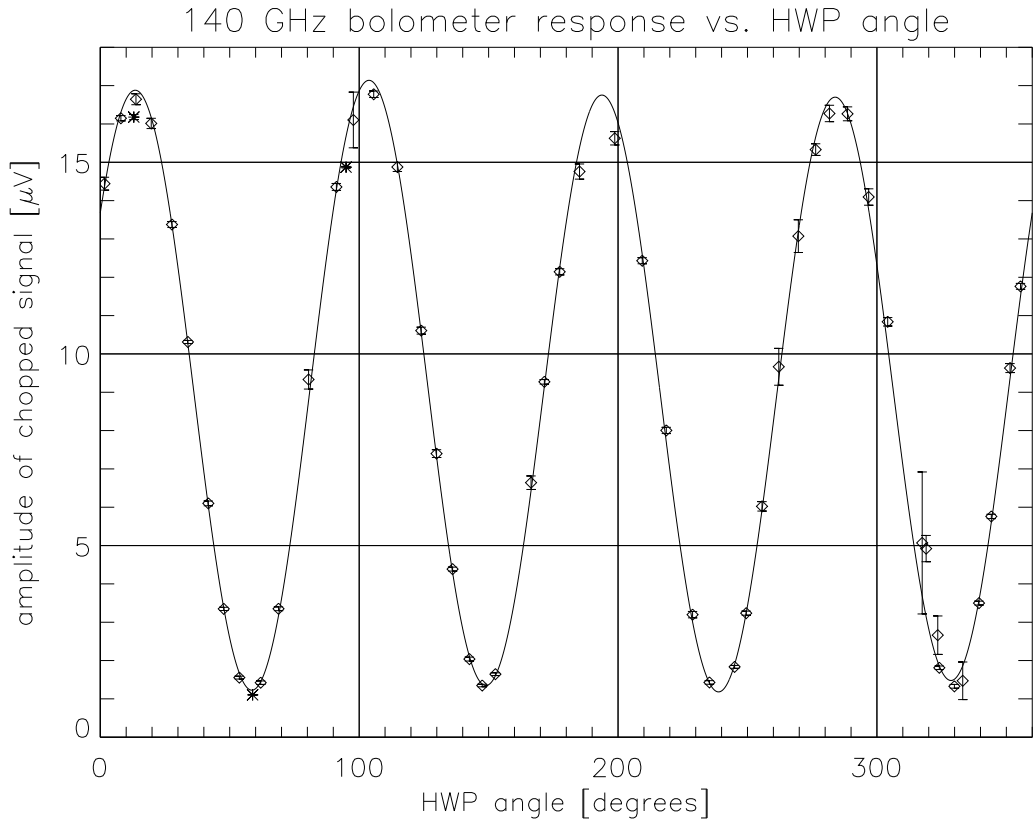,height=6.in,angle=90,clip=}}
\begin{center}
\parbox{6.5in}{
\caption{ \footnotesize A lab measurement of a polarized load.  The
setup for this measurement is described in
Section~\ref{sec:polarimeter}.  The solid curve plotted is the best
fit model comprised of $f_{0}$, $2f_{0}$, $3f_{0}$ and $4f_{0}$ sine
waves; the nine free parameters used in the fit include the amplitude
and phase of each sine wave and an overall offset. For this model, the
reduced $\chi^{2}$ = 1.22 for 42 degrees of freedom.  From the fit
parameters, we calculated the polarimeter efficiency to be 89\%.}
\label{fig:efficiency}
}
\end{center}
\end{figure*}

To ascertain the polarimeter efficiency, a polarized load was analyzed
in the lab before flight.  For this measurement, a wire-grid polarizer
was mounted on the cryostat window with its transmission axis oriented
parallel to that of the focal plane polarizer.  Thermal radiation from
a 273 K ice bath was chopped at $\sim$6.5 Hz with a 300 K aluminum
chopper blade covered with 0.25 inch thick Eccosorb LS-14 foam. The HWP
was then discretely stepped by hand in $\sim$5$^{\circ}$
intervals. Twenty seconds of data were collected at each HWP
orientation.  The amplitude of the chopped signal in the bolometer
time stream for one typical photometer was measured with a software
lock-in analysis and then plotted in Figure~\ref{fig:efficiency}.  A
nine parameter model consisting of sine waves for the first four
harmonics of $f_{0}$ was then fit to the data (solid curve); fit
parameters included the amplitudes and phases of each sine wave and an
overall offset. The level of polarization was then calculated from the
fit parameters using the standard definition $P = (A_{max} -
A_{min})/(A_{max} + A_{min})$ where $A$ is the amplitude of the
$4f_{o}$ signal.  This calibrated load was measured to be 86\%
polarized.  This corresponds to an overall polarimeter efficiency of
89\% which is in agreement with predictions that take into account the
HWP thickness, the known spectral response of the 140 GHz photometers, 
the convergence of rays at the aperature stop of the telescope and the
wire-grid polarizer efficiency.

Incident unpolarized light can become partially polarized inside the
instrument if it reflects off of the telescope mirrors at oblique
angles.  In addition, emission from the mirrors may also be partially
polarized.  To assess the level of instrumental polarization the
procedure outlined above was repeated with unpolarized light.  We
found the instrumental polarization to be 1\% for a typical 140 GHz
channel.

\subsection{Half-Wave Plate and Wire-Grid Polarizer}
\label{sec:hwp}

The 3.4 mm thick A-cut sapphire HWP was AR coated with a 0.013 inch
thick wafer of Herasil to maximize transmission.  The Herasil was
bonded to the sapphire with Eccobond 24, an unfilled, low viscosity
epoxy that was used to achieve glue layers as thin as 0.0005
inches. For MAXIPOL-0 we AR coated the HWP with a 0.010 inch thick
layer of Stycast 2850FT.  The switch from Stycast to Herasil was made
because Herasil thermally contracts in a way that is more compatible
with sapphire.

Since the AR coating was not birefringent, the two incident
polarization orientations had different coefficients of reflection;
this differential reflection gave rise to a rotation synchronous
signal at a frequency of $2f_{0}$.  To minimize this effect, we
calculated the AR coating thickness that would minimize the difference
in reflection coefficients given the spectral breadth of the 140 GHz
photometers, the thickness of the eccobond 24 layer and the oblique
incidence of rays.  Because the $2f_{0}$ signal resides out of the
polarization signal bandwith around $4f_0$ it is not a source of
systematic error (see Section~\ref{sec:systematics}).

The focal plane wire-grid polarizer, made by Buckbee-Mears, was
constructed from electroformed 0.0002 inch diameter gold wires bonded
to 0.0015 inch thick Mylar film at 250 lines per inch.  This flexible
material was mounted to a rigid ``roof-shaped'' frame that was
positioned over the horn openings.  This ``roof-shaped'' polarizer
reflected the unwanted polarization orientation out of the optical
path and into blocks of millimeter-wave absorbing material
\cite{bockblack} mounted on either side of the focal plane.  This
design reduced spurious signals due to reflections.

\subsection{HWP Drivetrain}
\label{sec:drivetrain}

\begin{figure*}[t]
\vspace{-0.85in}
\centerline{\epsfig{file=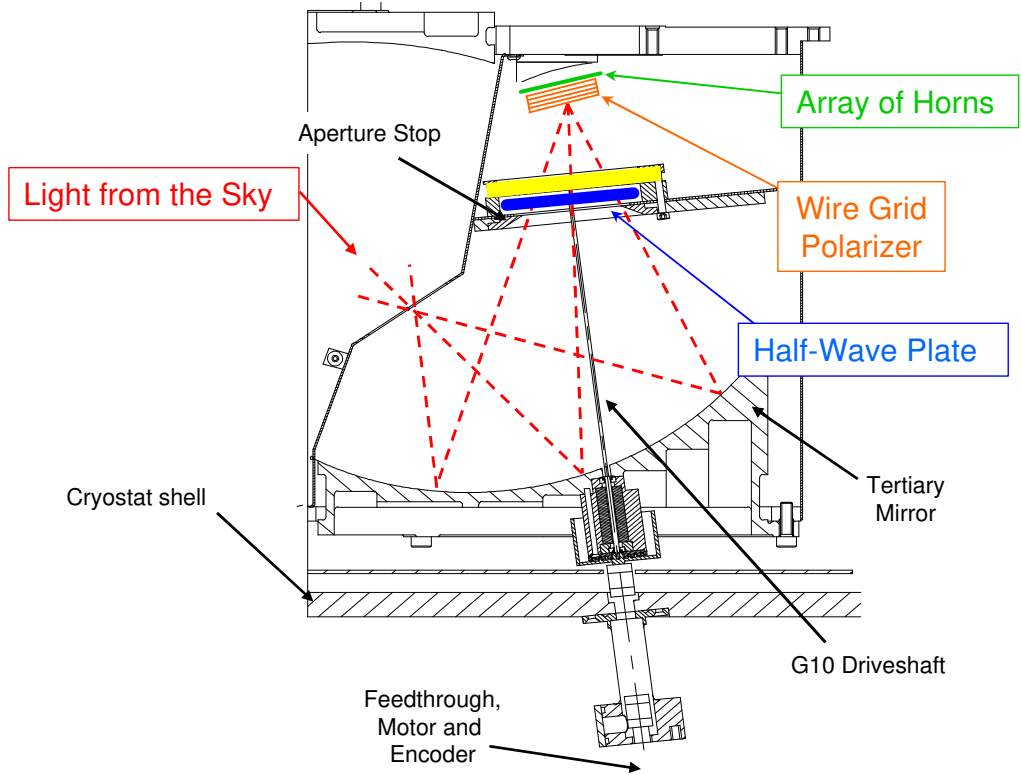,width=8.in,angle=0,clip=}}
\vspace{-0.75in}
\begin{center}
\parbox{6.5in}{
\caption{ \footnotesize A cross-sectional view of the MAXIPOL
polarimeter (Section~\ref{sec:drivetrain}).}
\label{fig:polarimeter}
}
\end{center}
\end{figure*}

The HWP rotated at $\sim$2 Hz during both MAXIPOL flights. This speed
was selected because it provided eight measurements of the of Stokes
parameters for one beam resolution element per scan period while
avoiding any significant $4f_{0}$ signal attenuation from the $\sim$10
ms bolometer time constant.  During operation, this rotation speed
proved to be vibrationally gentle; it did not excite any detectable
microphonic signals in the bolometer data.

The HWP was center turned near the Lyot stop of the telescope by a
0.078 inch diameter driveshaft (see Figure~\ref{fig:polarimeter}).  This
driveshaft penetrated the tertiary mirror and the cryostat shell and
was turned through a low-temperature ferrofluid rotary vacuum
feedthrough (Ferrofluidics FE51-122190A) by a feedback controlled
Kollmorgen U9M4 Servo Disc DC motor with high-altitude brushes mounted
outside the receiver. The orientation of the motor shaft, and
therefore the HWP, was measured with a 16-bit Gurley A25S optical
encoder.

The HWP was held in place near the Lyot stop with a Rulon-J sleeve
bearing embedded in a 0.4 inch thick disk of Zote foam. A sapphire
bearing was used in MAXIPOL-0 but was later found to exhibit less
favorable vibrational properties. This foam disk was mounted in the
optical path perpendicular to the chief ray. A polished,
hardened-steel sewing needle was passed through a center drilled hole
in the HWP; the assembly resembled a toy top.  One end of the needle
was slipped into the Rulon-J sleeve and the other end was rigidly
coupled to the driveshaft with 907 epoxy.

The driveshaft had three main parts.  Between the HWP and the tertiary
mirror, the driveshaft was made of thin wall G10 tubing (wall
thickness $\simeq$ 0.005 in).  The thin fiberglass G10 material
minimized both the thermal load on the HWP and the optical
cross-section of the exposed driveshaft, while retaining the desired
torsional driveshaft stiffness. A bearing assembly was mounted at the
back of the tertiary mirror to act as a thermal intercept and to
provided necessary mechanical stability for the driveshaft. The
bearing was made of Rulon-J (Vespel SP3) for MAXIPOL-1 (MAXIPOL-0).
Inside this bearing, the driveshaft was made of polished steel,
sputter coated with MoS$_{2}$; the shaft material was cromoly steel
(titanium nitride coated tungsten carbide) for MAXIPOL-1
(MAXIPOL-0). Between the tertiary mirror and the rotary vacuum
feedthrough at the cryostat shell the driveshaft was again made of
G10 tubing to minimize the thermal load on the HWP and the liquid
helium bath.

Laboratory testing was carried out to test the vibrational properties
of the drivetrain assembly at liquid helium temperatures. A mock-up of
the drivetrain was constructed and installed in a liquid helium
cryostat. With this setup, two bearing materials, Rulon-J and Vespel
SP3, were studied with microphones mounted near the bearings inside
the cryostat.  Rulon-J was chosen as the flight drivetrain bearing
material because of the low noise performance it exhibited over
several consecutive days of testing.


\section{MAXIPOL Flights}
\label{sec:flights}

\subsection{MAXIPOL-0}

During the $\sim$22 hour MAXIPOL-0 flight the NASA data transmitter
failed sporadically because of a broken solder connection.  As a
result, only a few $\sim$10 minute sections of bolometer data were
successfully recorded; we did not realize enough integration time for
CMB measurements.  The flight did provide us with the opportunity to
check the in-flight polarimeter performance and to test the new
daytime pointing camera, the sun shielding strategy and the HWP
driveshaft motor and encoder.

\subsection{MAXIPOL-1}

During the MAXIPOL-1 flight, we executed four different types of
telescope scans: a planet scan, a dipole scan, a CMB scan and a
foreground dust scan.

During the planet scan, the gondola yawed sinusoidally 2.5$^{\circ}$
peak-to-peak in azimuth at a slowly-rising elevation for $\sim$1 hour.
The scan period was 18 seconds.  During this time, Jupiter passed
through the field-of-view of the instrument and was detected by the
bolometers. This data set will be used to map the beam shape of each
photometer and calibrate the bolometer time streams given the known
millimeter-wave intensity of Jupiter.  We performed both daytime and
nighttime observations of Jupiter.

The CMB dipole was scanned by rotating the gondola 360$^{\circ}$ in
azimuth while holding the telescope at a constant elevation of
36$^{\circ}$ for 22 minutes.  A second $\sim$10 minute scan was
performed at an elevation of 50$^{\circ}$.  The period of a single
rotation was 18 seconds.  With this data set we will calibrate the
bolometer time streams given the large, known CMB dipole signal.

For the CMB and dust scans, the telescope tracked a guide star as it
swept across the sky. Simultaneously, the gondola yawed 2$^{\circ}$
peak-to-peak in azimuth with a period of 10 seconds. This telescope
motion combined with the inherent sky rotation produced bow tie
shaped maps.  To improve cross-linking, the telescope elevation
dithered periodically about the elevation of the guide star by $\pm$
0.2$^{\circ}$ with an elevation change ocurring every 10
minutes. Table~\ref{table:scans} summarizes the scan length and the
expected dust contribution for the five regions observed during
MAXIPOL-1.  Regions with strong dust contamination will be used to
characterize this foreground signal.

\begin{table*}[t]
\begin{center}
\begin{tabular}{ccccc}
&Observation Length&Average Dust&RMS Dust&Observation Time\\
&in [hours]&Level [$\mu K$]&Level [$\mu K$]&\\
\hline 
Beta Ursae Minoris&7.5&20&3.2&night\\
Polaris&2.25&173&23&day\\
Gamma Urase Majoris&2.5&11&2.5&day\\
Gamma Virgo&0.5&22&3.2&day\\
Arcturus&2.0&27&5.1&day\\
\hline
\end{tabular}
\parbox{6.5in}{
\caption{\footnotesize MAXIPOL-1 scan regions.  Five bow tie shaped
regions of the sky were mapped during MAXIPOL-1 by tracking five
different guide stars. The observation length and the expected dust
contribution for each region \cite{forecast} is presented.}
\label{table:scans}
}
\end{center}
\end{table*}

To monitor the bolometer temperature dependence of the calibration, a
fixed intensity millimeter-wave lamp mounted near the focal plane was
switched on for 10 seconds every 22 minutes.  The relationship
between the magnitude of the subsequent bolometer response and the
known bolometer temperature will be ascertained during data
analysis. The responsivity of each bolometer sample will be
interpolated given this relative calibration and the absolute Jupiter
and CMB dipole calibrations.


\section{Flight Performance}
\label{sec:systematics}

\begin{figure*}[t]
\begin{center}
\psfig{file=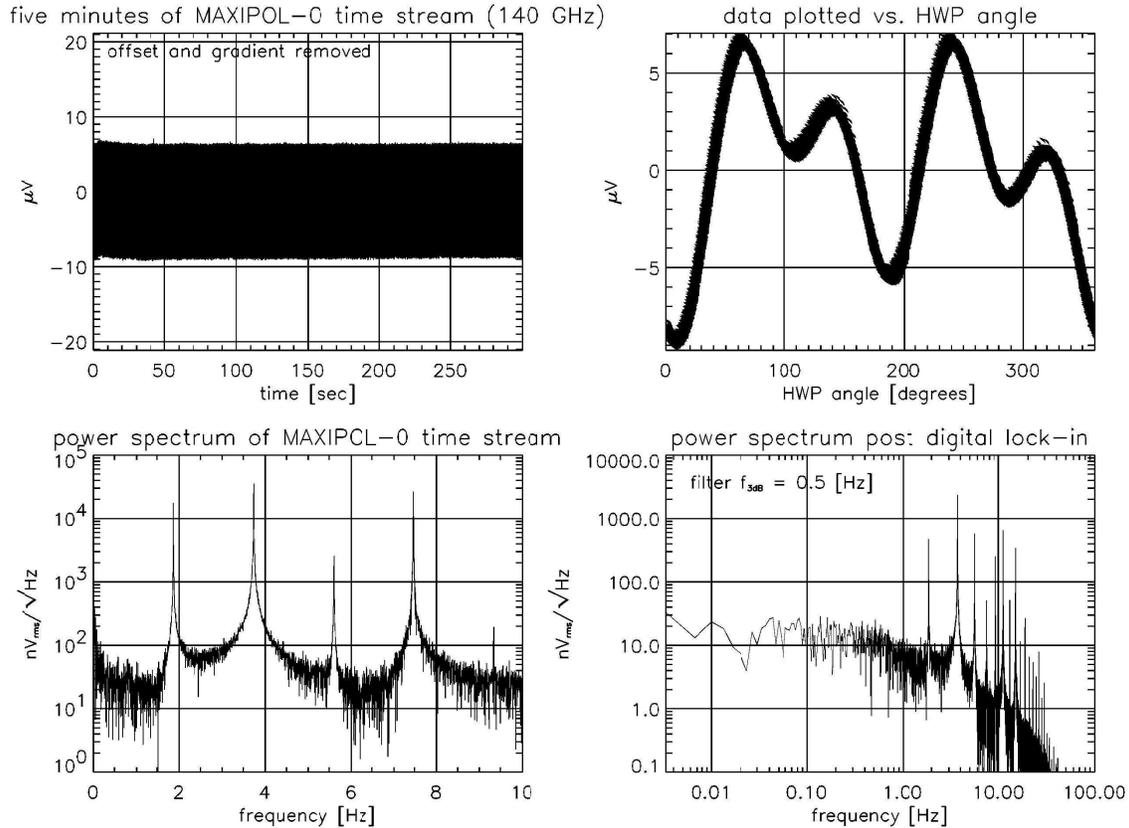,height=6.5in,angle=90,clip=}
\parbox{6.5in}{
\caption{ \footnotesize A software lock-in analysis of five minutes of
MAXIPOL-0 data (See the discussion in Section~\ref{sec:systematics}).}
\label{fig:data}
}
\end{center}
\end{figure*}

The purpose of this section is to demonstrate the viability of the
instrument. Five minutes of MAXIPOL-0 time stream data from one 140
GHz photometer are plotted in the upper left panel of
Figure~\ref{fig:data}. When this data is replotted versus HWP angle,
the systematic offsets become apparent (upper right panel).

Plotted in the lower left panel is the power spectrum of this data
set. The first peak appears at $f_{0}$ and the subsequent peaks appear
at the harmonics of $f_{0}$.  Laboratory measurements show that the
source of the $f_{0}$ signal is predominantly thermal emission from
the G10 drivetrain.  This thermal emission signal contributes at all
harmonics in the plot though it is subdominant in the $2f_{0}$ and
$4f_{0}$ peaks. The dominant $2f_{0}$ signal comes from the
differential reflection effect discussed in Section~\ref{sec:hwp}; the
dominant $4f_{0}$ signal comes from the instrumental polarization
signal discussed in Section~\ref{sec:polarimeter}.

As discussed in Section~\ref{sec:polarimeter}, polarization anisotropy
measurements will appear in the sidebands of the $4f_{0}$ signal.  If
the amplitude of the systematic offset at $4f_{0}$ varies, this
modulation may mimic polarization signals from the sky. To measure its
stability, we lock-in on the $4f_{0}$ offset with software using a
reference created from the HWP optical encoder data stream.  The time
dependence of the locked-in data provides a measure of the stability
of the amplitude of the $4f_{0}$ offset. The power spectrum of this
locked-in data is plotted in the lower right panel of
Figure~\ref{fig:data}. In the bandwidth of interest spanning 0 to 1 Hz
we recovered the nominal instrument noise level of $\sim$10
nV$_{rms}/\sqrt{\mbox{Hz}}$; this bandwith was set by the scan speed of the
telescope, the beam size and the angular size of the expected
structure on the sky.  We therefore conclude that variations in the
known systematic offsets will contribute less than the detector noise
to the the MAXIPOL data.

\section*{Acknowledgements}

We thank Danny Ball and the other staff members at NASA's National
Scientific Ballooning Facility in Ft. Sumner, New Mexico for their
outstanding support of the MAXIPOL program. We would also like to
thank the members of the Electronics \& Data Acquisition Unit in the
Faculty of Physics at the Weizmann Institute of Science in Rehovot,
Israel. MAXIPOL is supported by NASA Grants NAG5-12718, NAG5-3941 and
S-92548-F; a NASA GSRP Fellowship for B. Johnson; the Land McKnight
Professorship at the University of Minnesota, Twin Cities for
S. Hanany; and the Miller Institute at the University of California,
Berkeley for H. Tran.


\end{document}